\documentclass[journal]{IEEEtran} 
\IEEEoverridecommandlockouts
\usepackage{cite}
\usepackage{amsmath,amssymb,amsfonts}
\usepackage{algorithmic}
\usepackage{graphicx}
\usepackage{array}
\usepackage{epstopdf}
\usepackage{textcomp}
\usepackage{xcolor}
\usepackage{float}
\usepackage{caption}
\usepackage{subcaption}
\usepackage[normalem]{ulem}
\usepackage{siunitx}
\usepackage{booktabs}

\def\BibTeX{{\rm B\kern-.05em{\sc i\kern-.025em b}\kern-.08em
    T\kern-.1667em\lower.7ex\hbox{E}\kern-.125emX}}
\begin{document}

\title{TDC-less Direct Time-of-Flight Imaging Using Spiking Neural Networks\\
}

\author{
    \IEEEauthorblockN{Jack Iain MacLean$^1$}
    , \IEEEauthorblockN{Brian Stewart$^2$}
    , \IEEEauthorblockN{Istvan Gyongy$^1$}
    \\
    \IEEEauthorblockA{\textit{$^1$School of Engineering, The University of Edinburgh, Edinburgh, United Kingdom}}\\
    \IEEEauthorblockA{\textit{$^2$STMicroelectronics (R\&D) Ltd, Edinburgh, United Kingdom}}\\
    \IEEEauthorblockA{j.i.maclean@sms.ed.ac.uk}
}

\maketitle

\begin{abstract}
    3D depth sensors using single-photon avalanche diodes (SPADs) are becoming increasingly common in applications such as autonomous navigation and object detection. Recent designs implement on-chip histogramming time-to-digital converters (TDCs) to compress the photon timestamps and reduce the bottleneck in the read-out and processing of large volumes of photon data. However, the use of full histogramming with large SPAD arrays poses significant challenges due to the associated demands in silicon area and power consumption. We propose a TDC-less (and histogram-less) dToF sensor which uses Spiking Neural Networks (SNN) to process the SPAD events directly. The proposed SNN is trained and tested on synthetic SPAD events, and while it offers five times lower precision in depth prediction than a classic centre-of-mass (CoM) algorithm (applied to histograms of the events), it achieves similar Mean Absolute Error (MAE) with faster processing speeds and significantly lower power consumption is anticipated. 
\end{abstract}


\section{Introduction}
Direct time-of-flight (dToF) based LiDAR sensors are becoming increasingly widespread in applications requiring 3D imaging or proximity sensing, such as autonomous navigation or machine vision. Single-photon avalanche diodes (SPADs) are photodetectors that are seeing increased use in dToF imaging sensors as they provide single-photon sensitivity, low timing jitter, and are CMOS-compatible so arrays can be implemented with integrated digital processing. As SPAD-based sensors can generate large quantities of data, the bottleneck inherent in the data readout for off-chip processing is a significant challenge. The most common method for data compression is on-chip per-pixel histogramming of photon timestamps which takes advantage of state of the art 3D stacking technology to integrate SPAD and CMOS circuitry onto the same silicon die \cite{quantic_4x4} \cite{Istvan_on_chip} \cite{seo} \cite{chao_zhang}. Each photon timestamp is generated using using a time-to-digital converter (TDC) triggered on each SPAD event, with each timestamp used to increment the photon count in the corresponding histogram bin in memory. The histograms are accumulated over multiple laser cycles to improve the signal to noise ratio (SNR). The histogram memory can then be read out from the sensor and a peak estimation technique can be applied to extract detected surface depths, with some sensors performing the peak detection on-chip, such as in \cite{Istvan_on_chip}.\\
\indent Under high photon fluxes, photon pile-up can occur which prevents the detection of additional photons due to the dead-time inherent in the SPAD recharge method. Even at lower photon rates, TDC pile-up may occur, which distorts the histograms and prevents the detection of back-scattered photons at longer ranges \cite{gyongy_dtof}. One method to prevent this is the use of multi-event TDCs, such as those implemented in \cite{multi_tdc}, which allows for the recording of timestamps at a much higher rate.\\
\indent In recent years, neural network alternatives have been proposed to the conventional techniques of histogram processing (such as Gaussian curve fitting, continuous wavelet transform \cite{nhuyen_PD}, and center of mass filters \cite{center_of_mass}). These deep neural network implementations are capable of improving latency and reducing power consumption while matching or exceeding the accuracy of traditional techniques \cite{Chen_2022} \cite{Lidarnet} \cite{zang} \cite{SPADnet} \cite{histnet} \cite{german_sr}. However, the use of full histograms as input means that the associated silicon area (and power consumption) demands remain \cite{gyongy_dtof}. To circumvent the need to store and process large histograms on-chip, partial histogramming schemes \cite{partial_histogramming} \cite{Istvan_on_chip} have been developed, and, more recently a number of histogram compression schemes \cite{compressed_histogram} \cite{ingle2023countfree} have been proposed. The work presented in \cite{Spline_sketch_lidar}, for example, uses overlapping spline sketches of the detected photon events in order to compress the ToF data into a form of coarse histogramming. The advantage of this is that each photon arrival will increase the value in multiple bins by different amounts, allowing for a laser pulse duration shorter than the bin width without a significant loss in accuracy. A number of schemes have also developed methods which remove the need for histogramming entirely in favour of processing photon timestamps using photon timestamp averaging \cite{SPAD_linear}, traditional neural network architectures \cite{tommaso}, or the SPAD events themselves \cite{S.afshar}. \\
\indent SNNs operate asynchronously, encoding information as spikes allowing for better integration of time, phase, and frequency information. This grants the benefit of greater data sparsity, both spatially and temporally, while also allowing for relatively lower power consumption when compared to traditional synchronous neural networks \cite{SNN_summary}. Their main drawback, however, lies in the lack of a developed training method which puts limitations of the depth and complexity of the network \cite{TAHERKHANI2020253}. Despite this, SNNs have begun to see use in time-resolved imaging. There is an example of SNNs being used directly with SPADs in fluorescence-lifetime imaging microscopy, such as \cite{Lin_2024_WACV} which introduces two feed-forward SNN schemes to process the SPAD events achieving results comparable to a traditional recurrent Long Short-Term Memory (LSTM) architecture in simulations. To create deeper SNNs, a common technique is to convert conventional neural networks into a spiking equivalent following the methodology outlined in \cite{https://doi.org/10.48550/arxiv.1612.04052}. The conversion allows for traditionally trained networks to take advantage of the benefits SNNs provide, as in \cite{spike-spi}, with minimal degradation in accuracy. \\
\indent The work presented here investigates the application of SNNs on combined SPAD events directly for depth estimation, removing TDC and histogram memory requirement from the process. The network is based on the recursive Legendre Memory Units (LMU) architecture \cite{LMU}, due to its inherent ability to process long time series data with a modest number of parameters and internal state variables, achieving state of the art performance on learning temporal dependencies, lending itself to dToF applications. Training was carried out using NengoDL \cite{nengodl} in a Python environment. The model is trained on a synthetic dataset, an approach that has proved effective in previous studies on NN-based dToF processing \cite{Lidarnet}. Our dataset consists of randomised photon event streams generated in MATLAB \cite{MATLAB} following the optical model presented in \cite{ScholesStirling2023Fltd}. We compare the performance of the resultant network to a state of the art traditional CNN, LiDARNet \cite{Lidarnet}, and Centre of Mass (CoM) technique which use histograms created from the event data. The rest of the paper will comprise of the following three sections: Methodology II, Results III, and Conclusions IV.

\section{Methodology}
\subsection{Dataset}
\begin{figure}
\centerline{\includegraphics[width=0.5\textwidth]{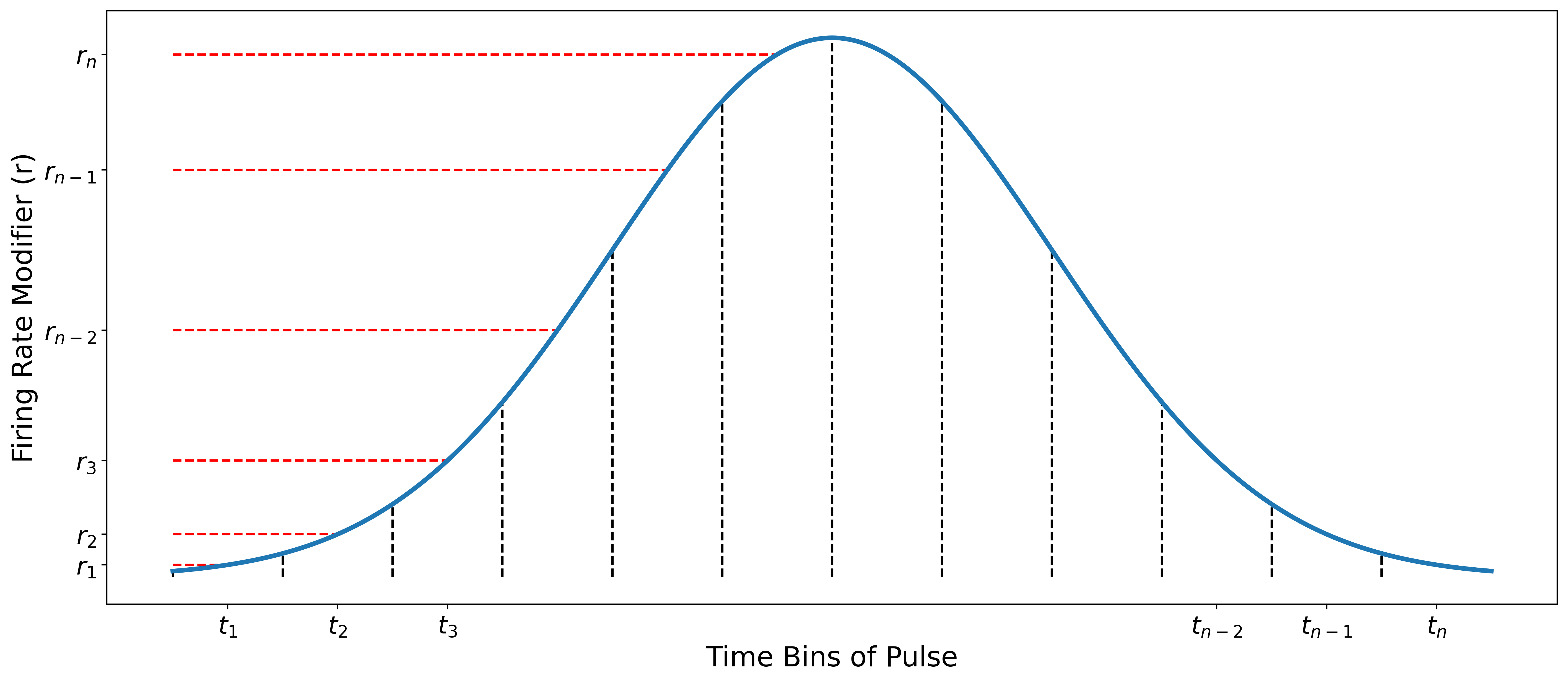}}
\caption{Gaussian profile rate estimation example\\}
\label{guass rate}
\end{figure}

\begin{figure}
\centerline{\includegraphics[width=0.5\textwidth]{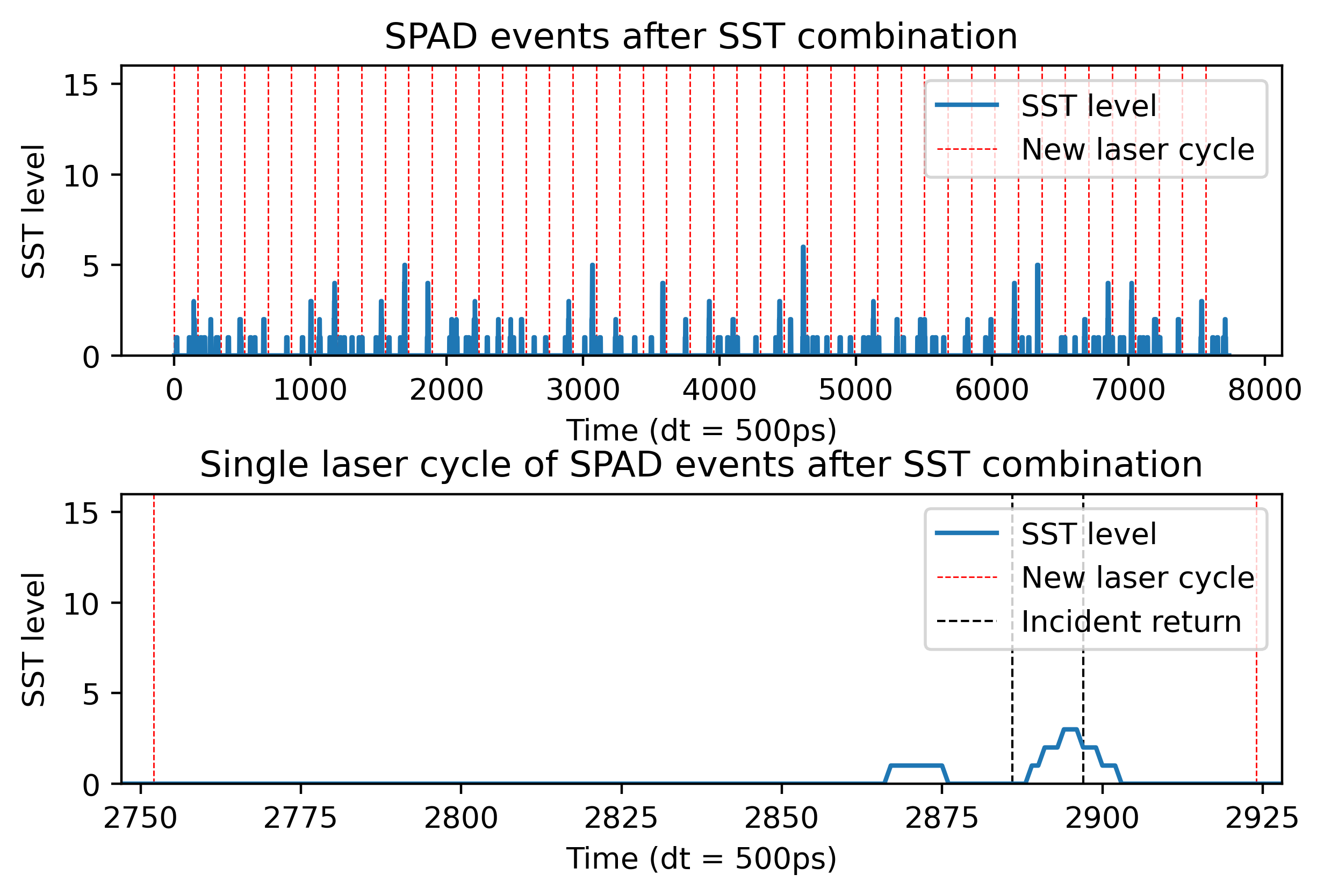}}
\caption{Example output from a single exposure of a 4$\times$4 SST combined SPAD pixel with a target distance of 10m, a reflectivity of 0.4, and an ambient level of 25 kLux. The lower graph provides a narrower time window covering a single laser cycle.}
\label{example waveform}
\end{figure}

The training dataset consists of 12,000 synthetic scene exposures, (10,000 samples for training, 2,000 for validation) with randomly generated depth, reflectivity, and ambient levels. The test dataset consists of four sets of 10,000 synthetic scene exposures at specific depths of 0.5~m to 10~m in steps of 0.5~m (500 exposures at each depth), with each set having a constant reflectivity (0.4 or 0.7) and ambient level (1 kLux or 25 kLux). The expected number of incident ambient and signal photon events per laser cycle within each exposure is calculated using the LIDAR model outlined in \cite{ScholesStirling2023Fltd}. Guided by SPAD modelling papers, \cite{TontiniAlessandro2020Nmos} \cite{SPAD_precision}, the number of laser cycles per exposure was taken to be 45, so that reasonable (cm) depth precision could be expected, whilst keeping the laser power at practical levels. The ambient photon arrival times are generated using a Poisson distribution across the whole exposure period. The arrival times of detected signal photons are also generated using a Poisson distribution. The return time window of the signal photons in each laser cycle is divided into multiple time bins with the base arrival rate being scaled by the temporal profile of the laser pulse, which we assume to be a Gaussian as illustrated in Figure \ref{guass rate} and outlined in Equation \ref{eq:0}, while the signal photon arrival rate is specified by Equation \ref{eq:1}.

\begin{equation} \label{eq:0}
    r_n = \frac{1}{\sigma \sqrt{2 \pi}}e^{\frac{-(t-\mu)^2}{2\sigma^2}},
\end{equation}
where $\sigma$ is the standard deviation of the temporal profile of the laser pulse and $\mu$ is the temporal position of the signal.

\begin{equation} \label{eq:1}
    \lambda = r_b r_n / T_{bin} N_{SPAD},
\end{equation}
where $\lambda$ is the chance for a photon to be detected at each timestep within the bin period, $r_b$ is the base signal photon arrival rate, $r_n$ is the scaling factor applied to the base rate to obtain the pulse profile, $T_{bin}$ is the length of the bin period, and $N_{SPAD}$ is the number of SPADs per macropixel being simulated. These generated photon events then trigger the simulated SPADs (with a certain amount of jitter) and the associated electronics, and are then combined using using a simulated implementation of the level sensitive SST combination technique \cite{SST}, which provides a higher dynamic range and is more resilient to photon pile-up compared to other combination methods such as an XOR tree. An example of the synthetic waveform used for training can be seen in Figure \ref{example waveform}. The network is also tested on an equivalent dataset which uses an asynchronous (clock-free) adder tree under the assumption of negligible delay between output bits. The histogram dataset used for the comparison network and CoM algorithm were created using an edge sensitive version of SST. Each histogram consists of 172 500ps bins which are accumulated using the SPAD event streams as their basis such that the data used to train and test both the SNN and comparison metrics are essentially the same. The synthetic datasets used in this work were written and generated in MATLAB \cite{MATLAB}. The full list of parameters used to emulate the desired LiDAR setup can be seen in Table \ref{datasetparam}.

\begin{table}[ht]
\caption{Synthetic SPAD event stream dataset parameters}
\begin{tabular}{ >{\raggedright\arraybackslash}m{7.5em}  m{1cm}  m{0.025cm}  m{2.5cm} m{1cm}}
\multicolumn{2}{l}{Illumination Source}                &  & \multicolumn{2}{l}{SPAD Sensor}                \\ \cline{1-2} \cline{4-5} 
Pulse Energy             & 640~nJ                      &  & SPADs per Pixel     & 16 \\
Pulse Duration (FWHM)    & 4~ns                  &  & Pixel Area          & 1600~$\mu m^2$ \\
Wavelength               & 940~nm                      &  & PDE                 & 18.5~\%  \\
Laser Cycles             & 45                          &  & SPAD Dead Time      & 4.3~ns \\
Beam Divergence          & 11.25$^\circ$               &  & SPAD Quench / Recharge Type& Passive \\
Laser Trigger Jitter     & 100~ps                      &  & SPAD Firing Jitter (FWHM)  & 100~ps   \\
                         &                             &  & Combination Tree    & SST or adder   \\
Optical System           &                             &  & SST Clock Period    & 500~ps   \\ \cline{1-2}
F number                 & 1.2                         &  &                     &           \\
Lens Transmittance       & 0.5                         &  &  \multicolumn{2}{l}{Additional Parameters} \\ \cline{4-5}
Optical Filter Bandwidth & 10~nm                       &  &  Simulation Timestep      & 1~ps \\
                         &                             &  &  Ambient Level    & 0$\rightarrow$30~kLux  \\
                         &                             &  &  Reflectivities  & 0.25$\rightarrow$1.0       
\end{tabular}
\label{datasetparam}
\end{table}

\subsection{Network Architecture}
\begin{figure}
\centerline{\includegraphics[width=0.4\textwidth]{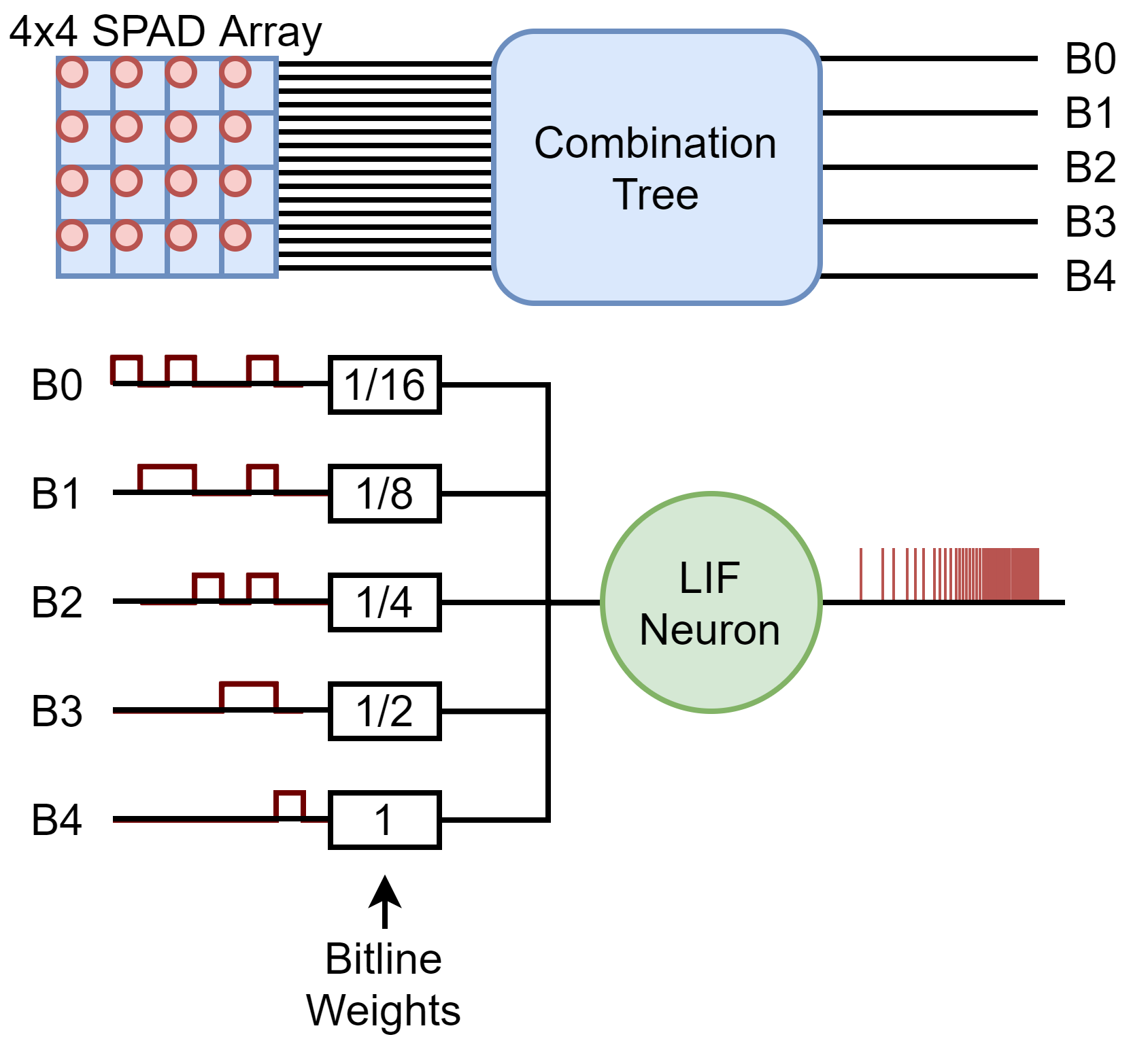}}
\caption{SPAD events to combined signal to spikes process}
\label{SST_conv}
\end{figure}

\begin{figure}
\centerline{\includegraphics[width=0.5\textwidth]{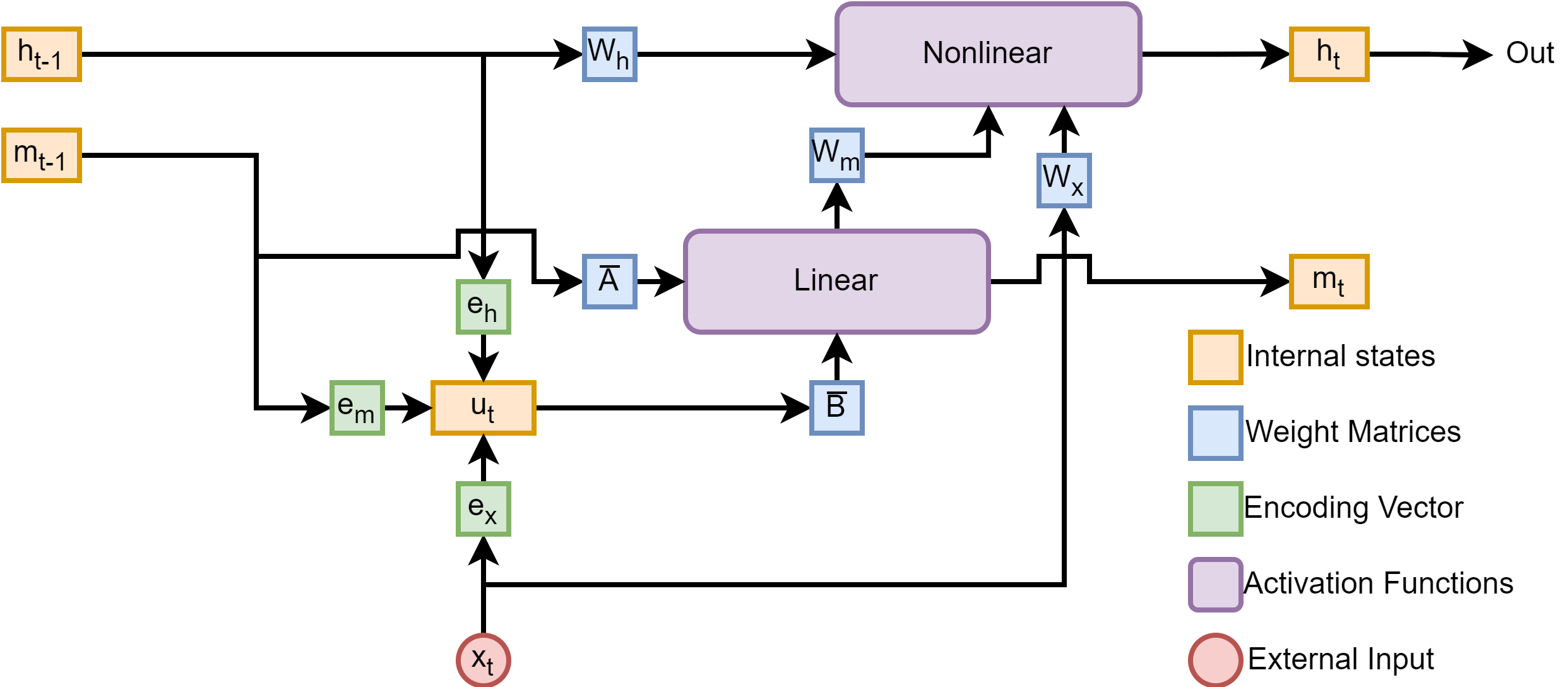}}
\caption{Internal structure of the LMU recurrent layer \cite{LMU}}
\label{LMU}
\end{figure}

\begin{figure}
\centerline{\includegraphics[width=0.5\textwidth]{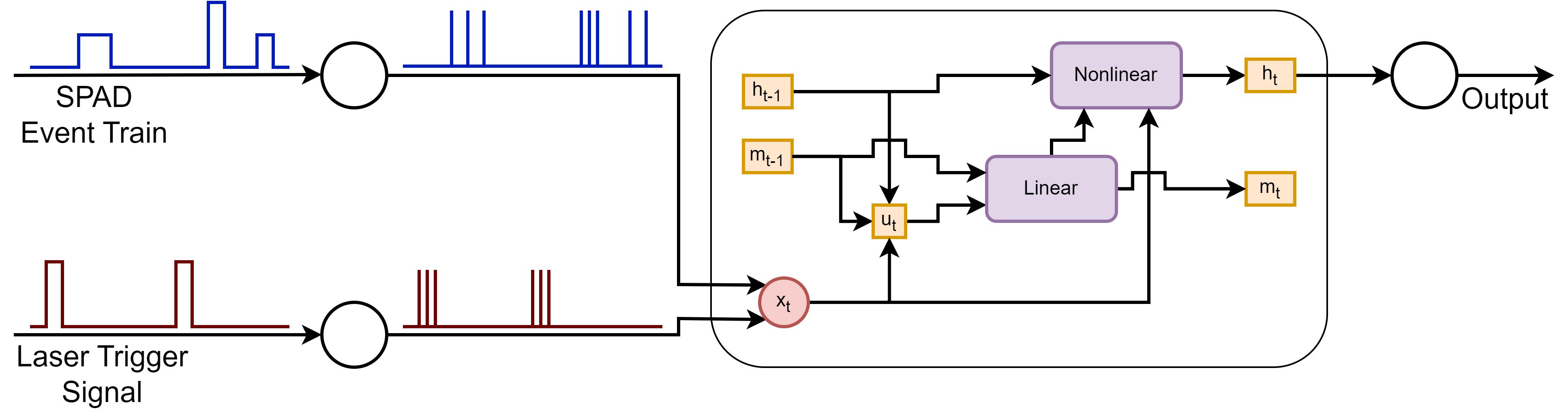}}
\caption{Overview of system architecture}
\label{SysArch}
\end{figure}

The proposed network utilises both the SPAD event stream and the laser trigger signal as inputs to the network. These signals are then converted into spike trains using Leaky Integrate-and-Fire (LIF) neurons \cite{koch1998methods} and fed to the next layer which comprises a Legendre Memory Unit (LMU) \cite{LMU}, with the SPAD event stream being converted as shown in Figure \ref{SST_conv}. The LMU is a recurrent layer which is capable of learning continuous time data with time steps approaching zero and minimal internal state-variables. It has been shown to outperform state of the art Recurrent Neural Networks (RNNs) on the permutated sequential MNIST dataset, and with less internal state variables \cite{LMU}. The internal structure of the LMU can be seen in Figure \ref{LMU}, and the governing equations are outlined in Equations \ref{eq:2}-\ref{eq:4}.
\begin{equation} \label{eq:2}
    u_t = e_x^T x_t + e_h^T h_{t-1} + e_m^T m_{t-1},
\end{equation}
\begin{equation} \label{eq:3}
    m_t = \Bar{A} m_{t-1} + \Bar{B} u_t,
\end{equation}
\begin{equation} \label{eq:4}
    h_t = f(W_x x_t + W_h h_{t-1} + W_m m_t),
\end{equation}
where $e_x$, $e_h$, and $e_m$ are learned encoding vectors. Matrices $\Bar{A}$ and $\Bar{B}$ are the discretised ideal state space matrices. $W_x$, $W_h$, and $W_m$ are learned kernels. In this implementation, the nonlinear function in Equation \ref{eq:3} is implemented using a LIF neuron which can accurately represent a ReLU activation function \cite{Cao2014SpikingDC}. The overall system architecture can be seen in Figure \ref{SysArch}. The proposed implementation of the LMU layer consists of 4 hidden units and 56 memory units, for a total of 60 internal state variables and 3,506 parameters. The ensemble of hidden units is then connected to the output layer which decodes the resultant spike train into a depth value. 

\subsection{Training Environment}
The network is simulated in Python using Nengo \cite{nengo} and trained using NengoDL \cite{nengodl}. Nengo is a tool which allows for the creation and simulation of SNNs based on the Neural Engineering Framework (NEF) \cite{NEF} while NengoDL is a tool which applies deep learning methods to the created networks to optimise their parameters and allows for the training of deeper multi-layered models. The network itself is trained using a LogCosh loss function and the Adam optimiser \cite{Adam}.

\section{Results}
\subsection{Synthetic Data}

\begin{figure}
\centerline{\includegraphics[ width=0.5\textwidth]{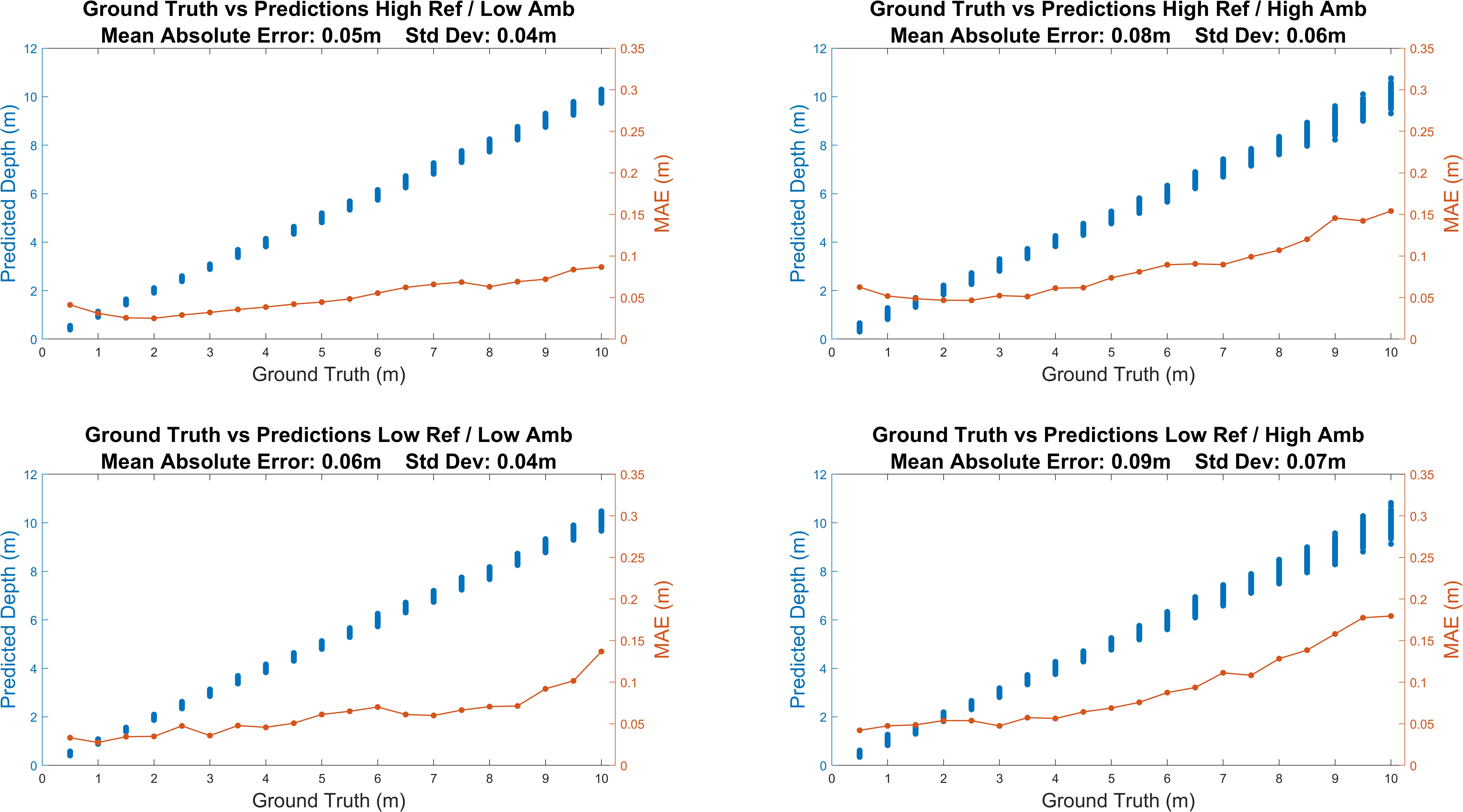}}
\caption{SNN prediction precision vs ground truth on test dataset}
\label{LMU_TestvsGT}
\end{figure}

\begin{figure}
\centerline{\includegraphics[width=0.5\textwidth]{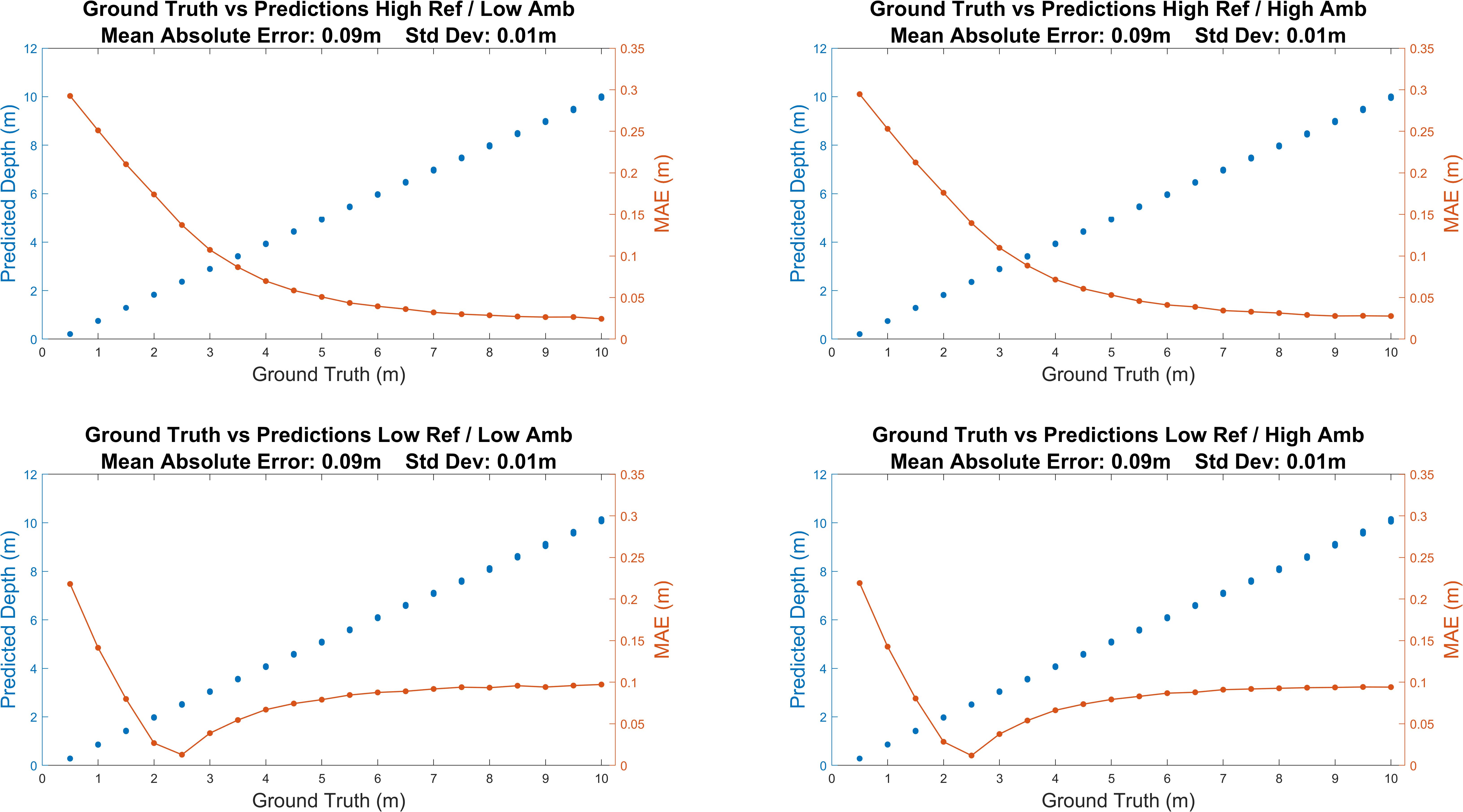}}
\caption{CoM prediction precision vs ground truth on test dataset}
\label{CoM_TestvsGT}
\end{figure}

\begin{figure}
\centerline{\includegraphics[width=0.5\textwidth]{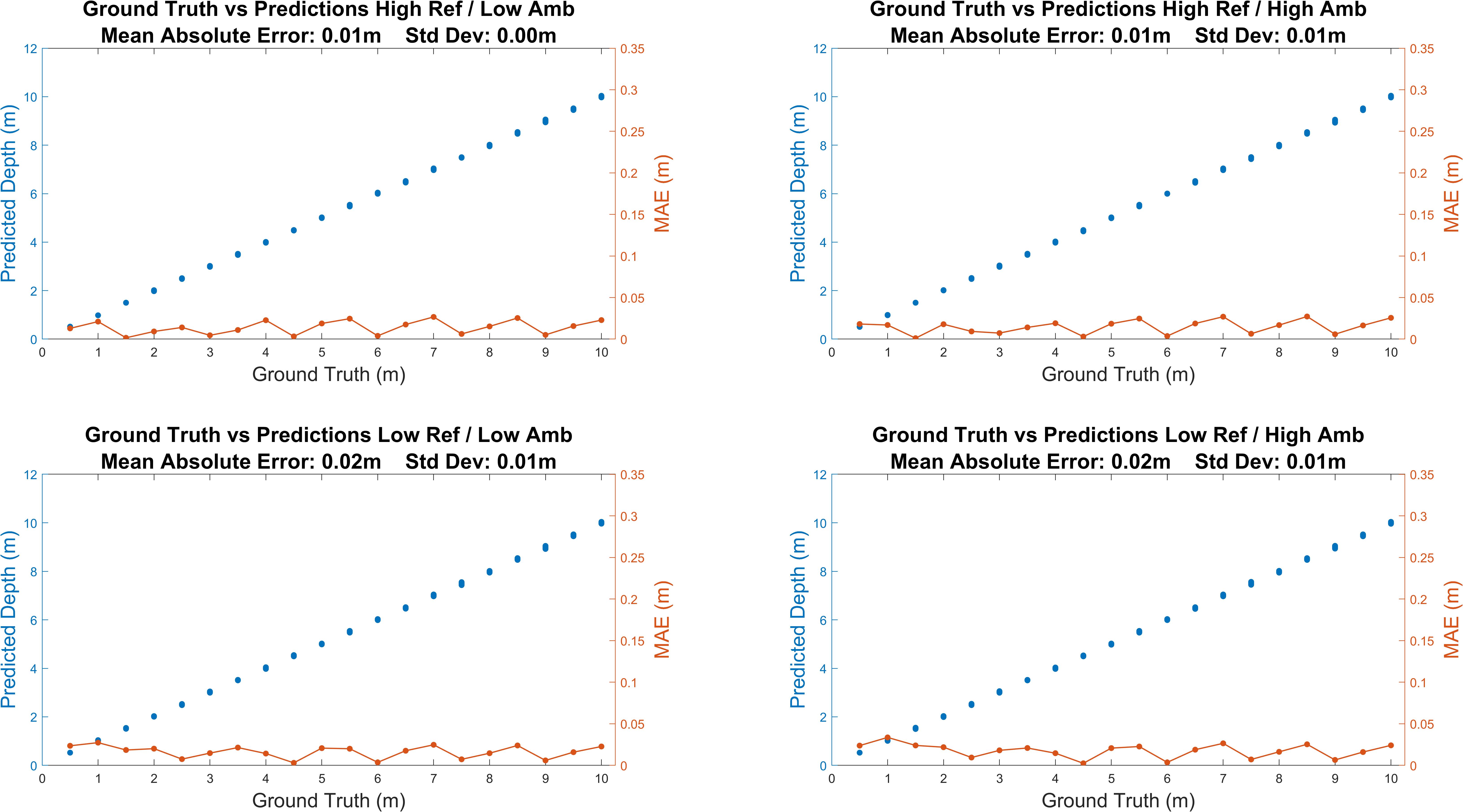}}
\caption{LiDARNet prediction precision vs ground truth on test dataset}
\label{Lidarnet_TestvsGT}
\end{figure}

The prediction results of the proposed network on the four synthetic test datasets can be seen in Figure \ref{LMU_TestvsGT}, and demonstrates the networks ability to predict the depths represented in the SPAD event stream. Under more challenging conditions, where SBR, defined as the number of incident signal photons versus background photons $\frac{N_{sig}}{N_{back}}$, is approximately 1 at 10m, the network is capable of predicting over 80\% of the surface depths within 15cm of ground truth. When tested on data generated using an asynchronous adder combination tree, without retraining the SNN, the degradation in MAE was found to be minimal ($<1$~cm) across the 4 different test scenarios. The expectation is that the network could be trained to perform even better in low SNR conditions, provided that the number of laser cycles per exposure was increased given the LMU architecture's ability to process long time series data. However, this was impractical to test given the already large data size testing the limits of available hardware (NVIDIA GeForce RTX 3090) and the long training times. 

\begin{table}[ht]
\caption{Comparison of this work with traditional peak detection and neural network approaches}
\begin{tabular}{ >{\raggedright\arraybackslash}m{7em}  m{1.5cm}  m{1.5cm}  m{1.5cm}}
                              & This Work   & CoM Peak Extraction   & LiDARNet Style CNN \\ \cline{2-4} 
Mean Absolute Error / std *$^1$& 0.07 (0.05)& 0.09 (0.01)           &  0.01 (0.01)       \\[0.1cm]
No. of Parameters             & 3,506       & N/A                   &  327,644           \\[0.1cm]
Estimated Power Consumption   & 204~pJ    & 59.52~nJ   *$^2$ *$^3$&  36.47~$\mu$J *$^2$ \\[0.1cm]  
Latency                       & 1 cycle     & 4 cycles              & $1.1\times10^5$ cycles
\end{tabular}
{\raggedright \\$^1$ Averaged across all four test datasets\\$^2$ This value only considers the processing of the histogram, in actuality this value would be higher due to the addition of the TDC and associated histogramming components. e.g. for a 16 bin histogram with 1 ns bin width this would increase energy consumption by 2.14 $\mu$J per pixel per inference \cite{Quantic_alternate}\\$^3$ Based on 8 bin \cite{Istvan_on_chip} and not 172 bin histogram \par}
\label{Comparison}
\end{table}

\indent The results of the proposed network are comparable with the Centre of Mass (CoM) technique applied to the histogram test data, Figure \ref{CoM_TestvsGT}, though is worse than the LiDARNet based CNN, Figure \ref{Lidarnet_TestvsGT}. The majority of the error in CoMs predictions over the dataset come from the close range distances where the SPADs are saturated due to the high photon flux, thus distorting the histograms and resulting in under prediction. While the CNN and SNNs data has the same saturation, they have both learned to recognise and adjust their predictions accordingly to avoid any under prediction. If this under prediction was ignored, then CoM has a similar MAE to the CNN.\\
\indent The main advantage of the SNN however lies in the power consumption, latency, and silicon area used to process each macropixel's SPAD events. One thing to note, however, is that the LiDARNet implementation here is not a hardware optimised implementation. The proposed network uses approximately three orders of magnitude less energy than CoM and six orders less than the CNN to make a single depth prediction on the presented data. This energy consumption is more significant when the energy needed for the TDC and the histogram memory is included (2.14 $\mu$J for a 16-bin 1 ns bin width histogram \cite{Quantic_alternate}). For latency, the SNN also has the advantage as it is processing the SPAD events as they occur so that it has a prediction result by the end of the exposure such that the only latency is the time taken to sample the output. The speed of CoM and the CNN are tied to the clock speed they are driven at. CoM has been successfully simulated operating at a 50MHz clock rate, so assuming both CoM and the CNN are driven at the same clock rate then CoM has a measured processing latency of 80ns for an 8 bin histogram while the CNN was calculated to have a 2.2ms latency for a 172 bin histogram. For the area required to process each exposure is difficult to determine for this work as the exact size and layout of the neurons has not yet been determined, however, one key advantage is that as the SNN does not require a TDC or histogramming components. This means that it is potentially saving over 7,525 $\mu m^2$ per macropixel for the histogram memory (based on scaling the area of 8, 12-bit/bin histogram on 40nm CMOS in \cite{Istvan_on_chip} to 172 bins) compared to CoM and the CNN.\\
\indent The SNN energy consumption was estimated based on the average number of spikes per exposure over the test dataset. An estimate was performed as in the absence of a detailed circuit diagram, which has not yet been performed, it is difficult to make an accurate prediction of the energy consumption. To that aim, we intend to provide a number of possible hardware configurations to give an indication of what the possible power consumption could be given select neuron architectures. The estimation of the energy consumption is calculated as a product of the number of spikes and the energy consumed to generate and propagate that spike, as in Equation \ref{eq:5}.

\begin{equation} \label{eq:5}
    E_{T} = E_{s} N_{s} + E_{n} N_{n}, 
\end{equation}
where $E_T$ is the total energy per exposure, $E_s$ and $E_n$ are the energy per synaptic and neural spike respectively, and $N_s$ and $N_n$ are the average total number of synaptic and neural spikes per exposure. In total, it was estimated that per exposure there were 15,000 neural spikes and 916,000 synaptic spikes. For the neural and synaptic spike energies, we use the energy consumption of 5 different suitable LIF neuron architectures \cite{BTBT_LIF} \cite{Graphene_LIF} \cite{benchmark_lif} \cite{L_BIMOS_LIF} \cite{LIF_TFET} to estimate the energy consumption for the neural spikes. There is however a lack of information regarding the energy consumption to propagate the spikes through the synapses, so this value has also been estimated based on the power ratio reported between neural and synaptic spikes seen in other SNN papers such as \cite{Graphene_LIF} \cite{Ara_shawkat} \cite{Loihi} which present factors of 8.67, 26.04, and 3.43 higher spike per neuron energy versus per synaptic spike respectively. Based on this, we have selected a ratio of 10 for a conservative estimate. The estimated energy consumption per exposure, based on the outlined methodology, for different configurations can be seen below in Table \ref{LMU_energy_estimation}. For comparison to the other sensors, we have selected the energy estimation based on \cite{BTBT_LIF}. We note that the power consumption of the final, ADC stage for converting the SNN's depth output to a digital code is assumed to be negligible (compared to the rest of the system), given the existence of ADC designs achieving $\approx1$ pJ/conversion \cite{ADC}. The total energy consumption of the SNN processing is therefore estimated as 204~pJ per exposure. 

\begin{table}[ht]
\caption{SNN energy consumption estimates based on select existing LIF neuron architectures}
\begin{tabular}{@{}
  S[table-format=1.0]
  S[table-format=3.2]
  S[table-format=3.0]
  S[table-format=3.0]
  S[table-format=3.0]
  S[table-format=3.0]
  @{}
}
\multicolumn{1}{c}{Source} & \multicolumn{1}{c}{$E_n$} & \multicolumn{1}{c}{$E_s$} & \multicolumn{1}{c}{$E_n N_n$} & \multicolumn{1}{c}{$E_s N_s$} & \multicolumn{1}{c}{Total} \\ 
\midrule
\cite{BTBT_LIF} & 1.90~fJ  & 190~aJ  & 29~pJ  & 174~pJ & 203~pJ               \\
\cite{Graphene_LIF}& 400.00~fJ& 30~fJ& 603~fJ & 3~pJ& 4~pJ                  \\
\cite{benchmark_lif}& 2.18~pJ& 218~fJ & 33~nJ  & 200~nJ  & 233~nJ               \\
\cite{L_BIMOS_LIF}  & 180.00~fJ   & 18~fJ  & 3~nJ   & 17~nJ  & 19~nJ                  \\
\cite{LIF_TFET}  & 1.50~aJ & 150~zJ & 23~fJ  & 137~fJ  & 160~fJ            
\end{tabular}
\label{LMU_energy_estimation}
\end{table}

\indent The values of the CoM algorithm for energy and latency are based on the measured values of the on-chip implementation in \cite{Istvan_on_chip}, while the energy consumption and latency of the CNN was estimated using the tool Zigzag \cite{ZigZag}. In this case the energy and latency of the CNN were estimated as if it had been implemented on an Edge TPU \cite{edge_tpu} chip.


\subsection{Real Data}
\begin{figure}
\centerline{\includegraphics[trim={0.5cm 3.5cm 0.5cm 3cm},clip, width=0.5\textwidth]{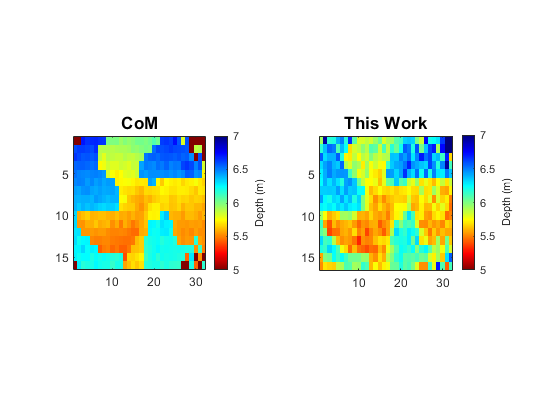}}
\caption{Comparison of depth maps generated by CoM and the presented work from real sensor data}
\label{real_depth}
\end{figure}

The presented SNN has also been tested on real data with results shown in Figure \ref{real_depth}. The real data was captured using the SPAD sensor \cite{192x128_robert} paired with a 300 ps pulse duration (FWHM) laser, $\approx$1 nJ pulse energy at a wavelength of 670nm. As the SPAD sensor produces binary time-stamp frames (rather than raw SPAD event stream required by the SNN), the data was reformatted into a form that better matched the synthetic data. 45 frames were thus combined into a single exposure to mimic the 45 laser cycles and multiple SPADs were grouped together into 4$\times$4 macropixels. This grouping of pixels also results in each macropixel capturing multiple surfaces (i.e. multiple peaks with the photons represented in histogram form) as opposed to the training data's one. Each frame's timestamps shifted in time based on the corresponding assumed laser cycle, and scaled by a factor of 3, so as the whole 3,870 ns input time range of the SNN would be used. The results are compared to a CoM applied to a histogram of the resultant SPAD events. The SNN can be seen to be much noisier than the CoM approach, but this expected given the differences between the training and real data (such as surface number, pulse duration, laser power, etc) with neural networks typically not performing well on data dissimilar to their training set. However, the network is still able to make reasonable predictions despite these differences, with the target object (a fan), still being visually recognisable.

\subsection{Comparison to Selected LiDAR Sensors}

\begin{table}[]
\caption{Per macropixel comparison of this work with other SPAD based LiDAR implementations.}
\fontsize{7.5}{9.5}\selectfont
\begin{tabular}{ >{\raggedright\arraybackslash} m{7em}  m{1.75cm}  m{1.75cm}  m{1.75cm}   }
                           & This Work    & \cite{quantic_4x4}     & \cite{Istvan_on_chip}  \\ \cline{2-4} 
Combination tree           & SST or adder         & XOR                    & OR                     \\
Data type                  & SPAD events  & Histogram              & Partial histogram      \\
On-chip depth              & Yes          & No                     & Yes                    \\
Processing technique       & Embedded SNN & CoM                    & Embedded CoM           \\
Technology                 & 20nm CMOS    & 40nm CMOS              & 40nm CMOS              \\
Power consumption @30 fps *$^1$ & 6.12~nW      & 64.36~$\mu$W*$^2$      & 1.79~$\mu$W*$^3$       \\
Area                       &              & 130/150~$\mu m^2$*$^4$ &                        \\
\end{tabular}

{\raggedright $^1$ This power estimate does not include the power consumption of the SPADs themselves\\$^2$ based on 131.8 mW for 1ns bin width and a 64x64 pixel array @ 30 fps\\$^3$ based on 12 mW for 63x32 pixel array @ 100 fps not including histogram generation\\$^4$ TDC area only \par}

\label{wider_Comparison}

\end{table}

\indent When compared to selected existing SPAD-based LiDAR sensors with in-pixel histogramming, see Table \ref{wider_Comparison}, the work presented has significantly lower power consumption when assuming each system is operating at 30 frames per second and scaled for a single pixel. However, this comes at the cost of higher errors in the depth measurements. Thus the scheme may be best suited to application where precise depth maps are not required but there are objects that need to be detected with low latency and power consumption.

\section{Conclusion}
In this work, a TDC-less LiDAR sensor has been simulated and shown to be able to process photon events at a lower power consumption per exposure and with a faster inference time than embedded CoM algorithms and CNNs, at the expense of lower precision. It has also been shown as robust to input data captured with a significantly different LiDAR set up than that with which it was trained. \\
\indent The expectation is that the presented SNN would be an ideal scheme to be implemented as a form of low power wake up for, say, automotive LiDAR, for example, to flag potential hazardous object in the vehicles path which could then be scrutinised by a more accurate but higher power traditional network. This would allow for the larger clocked system to be turned off until needed rather than constant polling of the scene. This will be the subject of further research as well as investigating a physical implementation, the effect of varying the clock speed of the SST combination tree, and the retraining of the network on more complex (e.g. multi-surface) synthetic data.

\bibliographystyle{ieeetr}
\bibliography{bibtex.bib}

\end{document}